# Generating photon-added states without adding a photon

S.U. Shringarpure and J.D. Franson
University of Maryland Baltimore County, Baltimore, MD 21250 USA

We show that a continuous range of nonclassical states of light can be generated using conditional measurements on the idler mode of an optical parametric amplifier. The output state is prepared by introducing a coherent state in the signal mode of the amplifier with a single photon in the idler mode, followed by a conditional measurement of a single photon in the output idler mode. By varying the gain of the amplifier, this approach can produce a coherent state, a photon-added state, a displaced number state, or a continuous range of other nonclassical states with intermediate properties. We note that this approach can generate a photon-added state even though the post-selected amplifier does not add any photons to the signal or idler modes. The ability to generate a continuous range of nonclassical states may have practical applications in quantum information processing.

### I. Introduction

There has been a large amount of research on the properties of various nonclassical states of light, such as photon-added coherent states [1-5] and displaced number states [6-8]. Nonclassical states of this kind are important resources for quantum information processing with continuous variables [9], quantum-key-distribution [10,11], boson sampling [12], quantum teleportation [13,14] and dense coding [15]. In this paper, we describe a method for preparing a continuous range of nonclassical states of light using post-selection on the idler mode of an optical parametric amplifier. By varying the gain of the amplifier, this approach can be used to generate a coherent state, a photon-added state, a displaced number state, and a continuous range of other nonclassical states with potentially useful properties.

The basic approach is illustrated in Fig. 1, where a coherent state is incident in the signal mode of an optical parametric amplifier with a single photon incident in the idler mode. We post-select the output state of the signal mode when a single photon is detected in the output idler mode. Since the signal and idler photons are emitted in pairs, the post-selection process ensures that no photons were emitted or absorbed in either mode. Nevertheless, the post-selection process can have the effect of creating a photon-added state for an appropriate choice of the gain. Other choices of the gain can produce a displaced number state or states that are orthogonal to a coherent state or a photon-added state, which may be useful for continuous-variable qubits.

Quantum state engineering methods to prepare various types of quantum states [16], such as photon-added/subtracted coherent states, thermal states, displaced number states [17], superpositions of number states [18], and truncated coherent states [19] have been explored using conditional measurements on beam splitters. These techniques have been very successful, but they have limited tunability of the prepared state due to the fixed transmittance of conventional beam splitters.

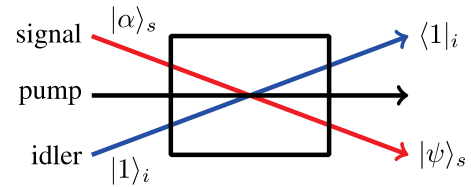

FIG. 1. Optical parametric amplifier with a coherent state in the input signal mode and a single photon number state in the idler mode. A measurement of a single photon in the output idler mode heralds the state of interest in the output signal mode. By varying the gain, this process can produce a continuous range of quantum states in the output.

The equivalence between a lossless beam splitter and an optical parametric amplifier when the input and output signals are appropriately interchanged has been previously discussed [20,21]. This equivalence allows an optical parametric amplifier to be used with conditional measurements in a way that is somewhat analogous to the use of a beam splitter. This has been applied, for example, to the noiseless attenuation of coherent states [22] and the preparation of various non-classical states [23-25], and it provides part of the motivation for the work reported here.

This paper is organized as follows. Section II derives the form of the post-selected output state and the corresponding probability of success. Sections III and IV examine the behavior of the final state as a function of the amplifier gain, including the special cases where the output is a displaced number state or a photon-added state. Section V describes the properties of the output in phase space using the Q-function for specific values of the gain. Section VI provides a summary and conclusions.

## II. State preparation

The time evolution operator $\hat{S}$ for an optical parametric amplifier can be written in a factored form given by [26,27]

$$\hat{S} = \frac{1}{g} e^{-\sqrt{g^2-1}\hat{a}^\dagger \hat{b}^\dagger/g} g^{-(\hat{a}^\dagger \hat{a} + \hat{b}^\dagger \hat{b})} e^{\sqrt{g^2-1}\hat{a}\hat{b}/g}. \quad (1)$$

Here $g = \cosh(\kappa t)$ is the gain, where $\kappa$ is the coupling strength of the amplifier and $t$ is the interaction time, while $\hat{a}$ and $\hat{b}$ are the annihilation operators of the signal and the idler modes respectively. Note that we can effectively tune the amplifier gain by varying the intensity of the pump. We will define $G \equiv \sqrt{g^2-1}/g$ for convenience.

We assume that a coherent state $|\alpha\rangle_s$ is introduced in the signal mode while a single photon number state $|1\rangle_i$ is incident in the idler mode of the amplifier. This corresponds to an input state of $|\alpha\rangle_s |1\rangle_i$, which is a simplified notation for $|\alpha\rangle_s \otimes |1\rangle_i$ in the tensor product space of the signal and idler modes. The transformation produced by the optical parametric amplifier is followed by a conditional measurement of a single photon in the idler mode, which can be represented by a projection operator $\hat{\Pi}$ given by

$$\hat{\Pi} = |1\rangle_i \langle 1|_i. \quad (2)$$

Thus, the final state $|\psi\rangle$ after these operations is given by

$$|\psi\rangle = \hat{\Pi}\hat{S}|\alpha\rangle_s |1\rangle_i. \quad (3)$$

Using a Taylor series expansion of the final exponential in Eq. (1) gives

$$e^{G\hat{a}\hat{b}}|\alpha\rangle_s |1\rangle_i = \left(1 + G\hat{a}\hat{b} + ...\right)|\alpha\rangle_s |1\rangle_i \\
= |\alpha\rangle_s \left(G\alpha|0\rangle_i + |1\rangle_i\right). \quad (4)$$

We see that only the first two terms of the expansion contribute after it acts on the input state. Similarly, the adjoint of the Taylor series expansion of the first exponential factor in Eq. (1) acting on the projection operator to the left also gives only two non-vanishing terms:

$$\hat{\Pi} e^{-G\hat{a}^\dagger \hat{b}^\dagger} = |1\rangle_i \langle 1|_i \left(1 - G\hat{a}^\dagger \hat{b}^\dagger + ...\right) \\
= |1\rangle_i \langle 1|_i - G\hat{a}^\dagger |1\rangle_i \langle 0|_i. \quad (5)$$

Next, we let the middle exponential factor in Eq. (1) act on the state obtained in Eq. (4). Expanding the coherent state in the number basis gives

$$g^{-(\hat{a}^\dagger \hat{a} + \hat{b}^\dagger \hat{b})} |\alpha\rangle_s \left(G\alpha|0\rangle_i + |1\rangle_i\right) \\
= \left(g^{-\hat{n}_s}|\alpha\rangle_s\right) g^{-\hat{n}_i}\left(G\alpha|0\rangle_i + |1\rangle_i\right) \\
= \left(e^{-|\alpha|^2/2} g^{-\hat{n}_s} \sum_{n=0}^{\infty} \frac{\alpha^n}{\sqrt{n!}}|n\rangle_s\right)\left(G\alpha|0\rangle_i + \frac{1}{g}|1\rangle_i\right) \\
= \left(e^{-|\alpha|^2/2} \sum_{n=0}^{\infty} \frac{1}{\sqrt{n!}}\left(\frac{\alpha}{g}\right)^n |n\rangle_s\right)\left(G\alpha|0\rangle_i + \frac{1}{g}|1\rangle_i\right) \\
= e^{-|G\alpha|^2/2} |\alpha/g\rangle_s \left(G\alpha|0\rangle_i + \frac{1}{g}|1\rangle_i\right). \quad (6)$$

Inserting Eqs. (1), (4), (5) and (6) in Eq. (3), we get the following post-selected (unnormalized) state in the signal mode

$$|\psi\rangle = e^{-|G\alpha|^2/2}\left(\frac{1}{g^2} - \frac{1}{g}G^2\alpha \hat{a}^\dagger\right)|\alpha/g\rangle_s. \quad (7)$$

Eq. (7) shows that the post-selected output state is a superposition of an attenuated coherent state and a photon-added coherent state, where the probability amplitudes of those two states can be controlled by varying the gain.

Eq. (7) can be rewritten in another useful form by using the fact that $\hat{a}|\alpha/g\rangle_s = g\hat{a}|\alpha/g\rangle_s$ which gives

$$|\psi\rangle = e^{-|G\alpha|^2/2}\left(\frac{1}{g^2} - G^2 \hat{a}^\dagger \hat{a}\right)|\alpha/g\rangle_s. \qquad (8)$$

Since $\hat{a}^\dagger \hat{a} = \hat{n}_s$, this gives the following expression for the final state:

$$|\psi\rangle = e^{-|G\alpha|^2/2}\left(\frac{1}{g^2} - G^2 \hat{n}_s\right)|\alpha/g\rangle_s. \qquad (9)$$

This form of the post-selected output state provides useful insight into the effects of the post-selection process as viewed in a number-state basis, as will be discussed in the next section.

The probability $P_s$ of success for the post-selection process is given by the norm of the final state $|\psi\rangle$ in Eq. (7), which can be shown to be

$$P_s = e^{-|G\alpha|^2}\left[\left(\frac{1}{g^2} - \left|\frac{\alpha}{g}\right|^2 G^2\right)^2 + \left|\frac{\alpha}{g}\right|^2 G^4\right]. \qquad (10)$$

The final state can then be normalized to give

$$|\psi\rangle = \frac{\left(\frac{1}{g^2} - G^2 \hat{n}\right)|\alpha/g\rangle}{\sqrt{\left(\frac{1}{g^2} - \left|\frac{\alpha}{g}\right|^2 G^2\right)^2 + \left|\frac{\alpha}{g}\right|^2 G^4}}. \qquad (11)$$

It can be seen that the probability of success is exponentially small for large values of $|G\alpha|$.

### III. Displaced photon number states

Eqs. (7) and (9) show that the post-selection process of Fig. 1 can be used to generate a continuous range of quantum states as we vary the gain. In this section, we will describe some of the properties of these states as a function of the gain. In particular, we will show that a specific value of the gain can be used to generate a displaced single-photon state.

Eq. (9) suggests that the value of the gain can be chosen in such a way that the coefficient $c_n$ in an expansion of the state in a basis of number states will vanish for a specific value of $n$. For example, an appropriate choice of the gain can cause $c_n$ to vanish when $n$ is equal to the mean photon number. In that case, the final state will have an asymmetric probability amplitude in the number state basis as illustrated in Fig. 2. This allows the final state $|\psi\rangle$ to be chosen to be orthogonal to the coherent state $|\alpha/g\rangle_s$, which may be useful in generating two orthogonal states for use as a qubit, for example.

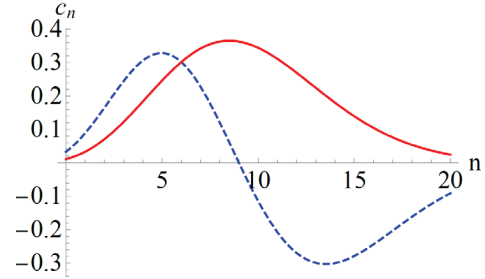

FIG. 2. Coefficients $c_n$ in the expansion of two states of interest in a basis of number states $|n\rangle$. The solid red line shows the coefficients $c_n$ for an attenuated coherent state $|\alpha/g_0\rangle$ for the case of $|\alpha|^2 = 10$, with $g_0$ given by Eq. (12). The dashed blue line shows the coefficients $c_n$ of the final post-selected state for a value of the gain $g = g_0$, which causes the final state to be orthogonal to the attenuated coherent state $|\alpha/g\rangle$. It can be seen that the cancellation of the two terms on the right-hand side of Eq. (9) gives $c_n = 0$ near the center of the distribution, so that the final state is approximately asymmetric about the mean photon number. The final state in this case is a displaced single photon state.

We will now show that a displaced number state can be produced by choosing a value of the gain given by

$$g = g_0 \equiv \frac{1}{\sqrt{1 - 1/|\alpha|^2}}. \qquad (12)$$

A displaced numbers state has the property that it is orthogonal to the corresponding coherent state, which may be a useful way to represent two orthogonal qubits. Inserting the value of the gain from Eq. (12) into Eq. (11) gives

$$|\psi\rangle_{g_0} = -\left(\frac{|\alpha|^2}{\alpha^*}\right)\left(\frac{\alpha^*}{g_0} - \hat{a}^\dagger\right)|\alpha/g_0\rangle_s. \qquad (13)$$

This state can be simplified by making use of the displacement operator $\hat{D}(\alpha)$ defined as usual by

$$\hat{D}(\alpha) = e^{\alpha \hat{a}^\dagger - \alpha^* \hat{a}}, \qquad (14)$$

which has the property that [7]

$$\hat{D}(\alpha)|0\rangle = |\alpha\rangle. \qquad (15)$$

The displacement operator satisfies the following commutation relationship with the photon creation operator [7]

$$[\hat{a}^\dagger, \hat{D}(\alpha)] = \alpha^* \hat{D}(\alpha). \qquad (16)$$

Eqs. (15) and (16) allows the $\hat{a}^\dagger |\alpha/g_0\rangle_s = \hat{a}^\dagger \hat{D}(\alpha/g_0)|0\rangle_s$ term in Eq. (13) to be rewritten in the opposite order of the operators. The commutator cancels the $\alpha/g_0$ term in Eq. (13), which gives

$$|\psi\rangle_{g_0} = -\left(\frac{|\alpha|}{\alpha^*}\right)\hat{D}(\alpha/g_0)|1\rangle_s. \qquad (17)$$

Eq. (17) shows that the post-selected amplifier produces a displaced number state as desired for this value of the gain.

Since $\hat{D}(\alpha)$ corresponds to a unitary transformation and $\langle 1|0\rangle = 0$, it can be seen that the displaced number state produced in this way is orthogonal to the corresponding coherent state $|\alpha/g_0\rangle_s$. The orthogonality of these two states can be understood from the asymmetric nature of the amplitudes $c_n$ in the photon number basis as shown in Fig. 2.

The displaced number state $|\psi\rangle_{g_0}$ also has the interesting property that it has the same average photon number as the initial coherent state in the input to the optical parametric amplifier. This can be shown by rearranging Eq. (12) into the form

$$|\alpha/g_0|^2 = |\alpha|^2 - 1. \qquad (18)$$

The average photon number for a displaced photon number state is given by [7]

$$\langle \hat{n}\rangle_{\hat{D}(\alpha')|n\rangle} = n + |\alpha'|^2. \qquad (19)$$

Combining Eqs. (18) and (19) for $n=1$ and $\alpha' = \alpha/g$, we see that the average photon number remains unchanged for $g = g_0$.

However, the variance in the photon number for the output state $|\psi\rangle_{g_0}$ is not the same as the input state. The variance for a displaced number states is given by [7]

$$\text{Var}(n)_{\hat{D}(\alpha')|n\rangle} = (2n+1)|\alpha'|^2. \qquad (20)$$

Combining Eq. (18) and Eq. (20) gives a photon-number variance of $3|\alpha/g_0|^2 = 3|\alpha|^2 - 3$.

We have shown that a gain of $g = g_0$ gives a displaced number state with $n=1$. We will now consider an arbitrary value of the gain and show that there is no contribution from displaced photon number states with photon number greater than 1. Using Eq. (16) in Eq. (11) allows the final state to be written in the form

$$|\psi\rangle = \frac{1}{\sqrt{N}}\left[\left(\frac{1}{g^2} - \left|\frac{\alpha}{g}\right|^2 G^2\right)\left|\frac{\alpha}{g}\right\rangle - \frac{\alpha}{g}G^2\left|\frac{\alpha}{g},1\right\rangle\right], \qquad (21)$$

where we have used the notation $\hat{D}(\alpha)|n\rangle = |\alpha, n\rangle$ and

$$N \equiv \left(\frac{1}{g^2} - \left|\frac{\alpha}{g}\right|^2 G^2\right)^2 + \left|\frac{\alpha}{g}\right|^2 G^4. \qquad (22)$$

Eq. (21) shows that the final signal state completely lies in the subspace of only two orthogonal states – the attenuated coherent state $|\alpha/g\rangle$ and the corresponding displaced single photon state.

**IV. Photon added states**

We showed in the previous section that the post-selection process can produce a displaced number state that is orthogonal to a coherent state. We now show that the post-selection process can also produce a photon added state in the limit of large gain.

This can be seen intuitively from Eq. (7), where the $G^2/g$ term becomes much larger than the $1/g^2$ term in the limit of large gain. As a result, the first term can be neglected in that limit and the second term gives a photon added state proportional to $\hat{a}^\dagger |\alpha/g\rangle$. Figure 3 shows a plot of the absolute value squared of the inner product between the final state and a single-photon added coherent state. The inner product approaches unity in the limit of large gain, which shows that the post-selected amplifier can generate a photon-added state in that limit as expected.

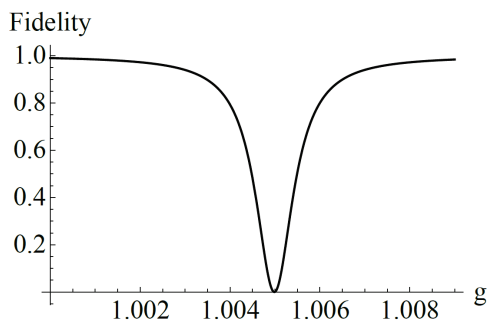

FIG. 3. A plot of the magnitude squared of the inner product of the output state with a photon-added state proportional to $\hat{a}^\dagger |\alpha/g\rangle$. The inner product is plotted as a function of the gain $g$ with $\alpha = 10$ ($n = 100$). It can be seen that the output state approaches a photon-added state in the limit of large gain, and that there is a gain $g = g_1$ where the output is orthogonal to a photon-added state.

It can also be seen from Fig. 3 that the inner product vanishes for a specific value of the gain $g = g_1$ where

$$\langle \alpha/g_1 | \hat{a} | \psi \rangle = 0. \qquad (23)$$

We can determine the value of $g_1$ by combining Eq. (23) with the requirement that the gain be real and greater than or equal to 1. It can be shown that this occurs for

$$g = g_1 \equiv \sqrt{\frac{2 - |\alpha|^2 + \sqrt{|\alpha|^4 + 4}}{2}}. \qquad (24)$$

Thus, the final state is orthogonal to the photon added state proportional to $\hat{a}^\dagger |\alpha/g\rangle$ for this value of the gain. The orthogonality of these two states may also be useful for generating continuous-variable qubits with two orthogonal states.

The contributions to the final state from an attenuated coherent state, a photon added state, and a displaced number state are summarized in Fig. 4, where the square of the projection of the final state onto these states is plotted over a relatively large range of the gain. It can be seen that there are values of the gain where the output state is purely a coherent state, a displaced number state, or a photon added state. In addition, it can be seen that the output is orthogonal to an attenuated coherent state or a photon added state for values of the gain equal to $g_0$ and $g_1$, respectively.

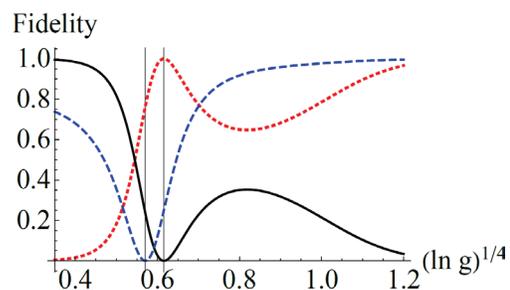

FIG. 4. Magnitude squared of the projection of the final state onto specific quantum states with $\alpha = 2$. The solid black line shows the projection onto an attenuated coherent state, while the dashed blue line shows the projection onto a photon-added coherent state. The dotted red line shows the projection onto a displaced photon number state. Values of the gain $g$ where the final state is orthogonal to an attenuated coherent state or a photon added state can also be seen. A logarithmic scale for the gain has been used in order to illustrate all of the relevant features.

## V. Q-functions

The Husimi-Kano Q-function [28,29] provides a convenient tool for visualizing the properties of quantum states as well as for calculating the expectation value of observables. For a single mode of the field, the Q-function $Q(\alpha)$ is defined as

$$Q(\alpha) = \frac{1}{\pi} \langle \alpha | \hat{\rho} | \alpha \rangle, \qquad (25)$$

where $\hat{\rho}$ is the density operator of the state. The Q-function corresponds to the diagonal matrix elements of the density operator in a basis of coherent states. For pure states, which we have here, this becomes

$$Q(\alpha) = \frac{1}{\pi}|\langle \alpha | \psi \rangle|^2, \quad (26)$$

where $|\psi\rangle$ is given by Eq. (21).

Figure 5 shows a plot of the Q-function for some specific values of the amplifier gain. Fig. 5(a) shows the Q-function for the initial coherent state $|\alpha\rangle$ corresponding to a gain of unity, while Fig. 5(b) shows a state that is orthogonal to a photon added state at a gain of $g = g_1 = 1.111$. The Q-function for a displaced number state that occurs at a gain of $g = g_0 = 1.154$ is shown in Fig. 5(c), and an arbitrary state at a higher gain of $g = 1.195$ is shown in Fig. 5(d); the output state approaches a photon-added state in the limit of large gain. All of these plots correspond to a coherent state amplitude of $\alpha = 2$, but similar results are obtained for other values of $\alpha$.

Fig. 5 exhibits the wide range of quantum states that can be produced using the post-selected amplifier illustrated in Fig. 1. It can be shown from Eq. (9) that cancellation between the $1/g^2$ and $G^2$ terms will cause the probability amplitude $c_n$ in a number-state basis to vanish at a particular value $n_0$ given by

$$n_0 = \frac{1}{g^2 G^2}. \quad (27)$$

It can also be shown that the Q-function vanishes at a coherent state amplitude $\alpha^*$ given by

$$\alpha^* = \frac{n_0}{(\beta/g)}. \quad (28)$$

The zero in the Q-function indicates that the state is orthogonal to the corresponding coherent state $|\alpha^*\rangle$. As the gain is increased for a fixed input signal amplitude, the zero can be seen to move inwards from infinity towards the origin. This can be viewed as a generalization of the orthogonality of the displaced number state of Eq. (17) to the coherent state $|\alpha/g_0\rangle_s$. It is consistent with the fact that, in the limit of large gain, the state becomes a single photon state that is orthogonal to the vacuum state. Roughly speaking, the existence of the zero in the Q-function corresponds to the orthogonality of the asymmetric $c_n$ coefficient in Fig. 2 to a particular coherent state.

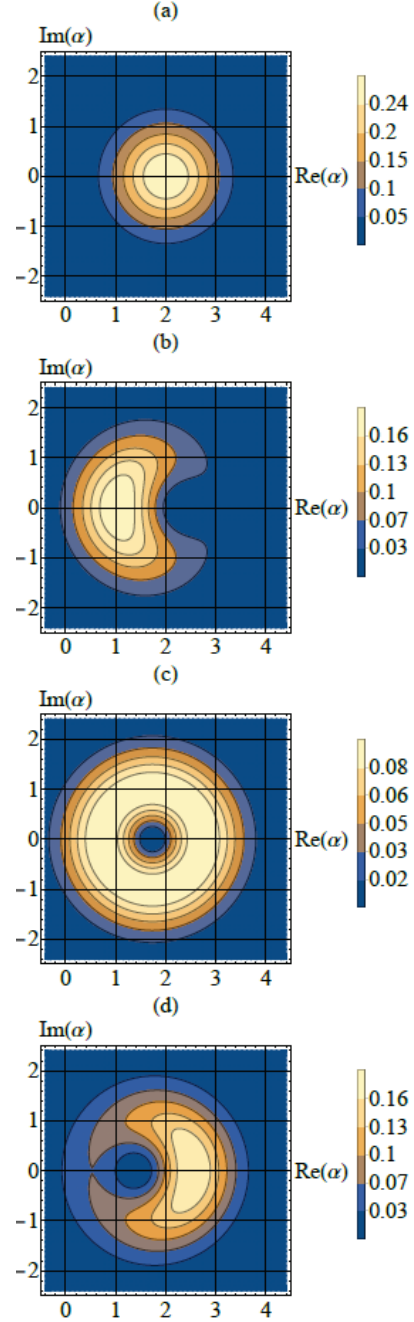

FIG. 5. Contour plots of the Q-function of the output state for an initial amplitude of $\alpha = 2$ for various amplifier gain values. (a) $g = 1$. This corresponds to the input coherent state. (b) $g = g_1 = 1.111$. This state is orthogonal to a photon added coherent state with an amplitude attenuated to $\alpha/g_1$. (c) $g = g_0 = 1.154$. This is a displaced single photon state with the amplitude of displacement attenuated to $\alpha/g_0$. (d) $g = 1.195$. This is an arbitrary state with a larger gain, which approaches a photon-added state in the limit of large gain.

## VI. Experimental Considerations

In any practical implementation of this approach, it will be essential to consider the effects of experimental errors such as photon loss, detector dark counts, and limited detector efficiencies. In order to analyze the effects of these experimental errors, we will consider the specific implementation shown in Fig. 6.

It will be assumed that the single photon input to the idler mode of the optical parametric amplifier in Fig. 1 is generated in one of the two output modes of a spontaneous parametric down conversion crystal (SPDC) as shown in Fig. 6. The presence of the single photon is heralded by post-selecting on a detection event in the other output path of the SPDC. We will assume that the spontaneous parametric down conversion process generates only a single pair of entangled photons at a time, which is a good approximation when the intensity of the laser used as a pump for the nonlinear crystal is sufficiently low. With this assumption, the only error in heralding a single photon is due to the dark counts in detector D1.

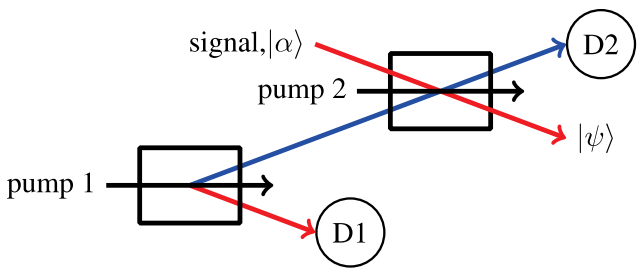

FIG. 6. Possible experimental setup for the generation of a continuous range of quantum states. The single photon required for input to the idler mode of an optical parametric amplifier (OPA) is generated by spontaneous parametric down-conversion (SPDC) in the lower nonlinear crystal which pumped by a laser. The detection of a single photon in detector D1 located in one of the output modes of the SPDC crystal heralds the presence of a single photon in the other mode. The single photon then enters an second nonlinear crystal used as an OPA, which is pumped by a second laser. The gain of the OPA can be varied by controlling the intensity of the second laser, which allows a continuous range of quantum states to be generated.

Post-selecting on the case in which a single photon is present in the output of the idler mode of the optical parametric amplifier in Fig. 1 will require a number resolving detector D2 as shown in Fig. 6. Here a dark count in detector D2 will produce an error in which it is assumed that an idler photon was present even when there were none. In addition, limited detection efficiency in detector D2 can produce an error in which it was concluded that one idler photon was present even though there were actually two or more. Errors of that kind are equivalent to photon loss combined with a perfect detection efficiency.

We will assume that there is negligible probability that higher photon number states ($n \geq 3$) will falsely indicate that a single photon was present in detector D2, which is a good approximation for relatively high detection efficiencies. We will also assume that there is negligible probability of having a dark count and a photon loss simultaneously. With these assumptions, we will denote the dark count probability in both detectors by $d$ and the probability that a photon count is lost (due to detector inefficiency or actual photon loss) by a probability $l$.

The fidelity $F$ of the final mixed state $\hat{\rho}$ with the ideal output state $|\psi\rangle$ of Eq. (9) is defined as

$$F = \langle \psi | \hat{\rho} | \psi \rangle. \tag{29}$$

$F$ can be written as a sum of terms corresponding to the inner products of $|\psi\rangle$ with the various states that are actually present in the output when a dark count or photon loss occurs. There are six different possibilities that could contribute significantly to an outcome in which both detectors appear to register a single photon. These outcomes will be labelled (0, 0), (0, 1), (0, 2), (1, 0), (1, 1) and (1, 2), where the first and second entries denote the actual photon number in the input and output modes of the amplifier respectively. For example, (0, 0) is an event in which two dark counts occurred, while (1, 1) corresponds to the case in which both detectors functioned correctly.

All six of these states can be calculated using techniques similar to those described above. The outcome (0, 0) corresponds to the noiseless attenuation of an input coherent signal state $|\alpha\rangle$ to $|\alpha/g\rangle$ as shown in Ref. [22]. The state (0, 1) corresponds to photon addition on $|\alpha/g\rangle$ while (1, 0) corresponds to $|\alpha/g\rangle$ itself. All six of these states can contribute to the fidelity in general since none of them are orthogonal to $|\psi\rangle$ for an arbitrary value of the gain.

The states (0, 2) and (1, 2) can also be calculated using similar techniques but they have a relatively complicated form. For simplicity, we can calculate a lower bound on the fidelity by assuming that the states (0, 2) and (1, 2) are approximately orthogonal to the desired state $|\psi\rangle$. In that case, the lower bound on the fidelity is given by

$$F = d^2 |\langle \phi_{(0,0)} | \psi \rangle|^2 + d(1-d-l) |\langle \phi_{(0,1)} | \psi \rangle|^2 \\ +(1-d)d |\langle \phi_{(1,0)} | \psi \rangle|^2 + (1-d)(1-d-l) |\langle \phi_{(1,1)} | \psi \rangle|^2. \tag{30}$$

Figure 7 shows the lower bound on the fidelity as a function of $d$ for several values of $l$. We have also plotted the actual fidelity including the contributions from the states $(0, 2)$ and $(1, 2)$. It can be seen that limited detection efficiency and dark counts can both have an effect on the fidelity.

The detectors used in pulsed down-conversion experiments have dark counts corresponding to $d$ as low as $10^{-6}$, so that dark counts should have a relatively small impact on the fidelity. Superconducting detectors can have efficiencies of about 95% ($l \leq 0.05$) which would also have a minimal effect on the fidelity, whereas the more common silicon photo avalanche diodes can have detector efficiencies up to ~74% [30]. In both cases, the dominant error source is more likely to be actual photon loss due to coupling between fibers or absorption in filters.

### VII. Summary and conclusions

We have shown that post-selection on the idler mode of an optical parametric amplifier can generate a continuous range of quantum states with different properties. As illustrated in Fig. 1, a coherent state is assumed to be incident in the signal mode while a single photon is incident in the idler mode. Post-selection on a single photon emerging in the idler mode gives an output state whose properties depend on the gain of the amplifier. The states that can be generated in this way include a coherent state, a displaced number state, and a photon added state, along with a continuous range of states with intermediate properties.

One of the interesting features of this approach is that no photons are absorbed or emitted in the idler mode due to the post-selection process, and no photons are absorbed or emitted in the signal mode either since the photons are only absorbed or emitted in pairs. As a result, one might suspect that the amplifier has done nothing. Nevertheless, the post-selection process can change the probability amplitudes $c_n$ of the state in a number-state basis, since different values of $n$ will have different probability amplitudes for producing the post-selected output. In that respect, these results are somewhat similar to an earlier paper [31] in which we considered post-selecting on an ensemble of absorbing atoms, accepting only those events in which the atoms remained in their ground states. Although the atoms may appear to have done nothing, the post-selection process can increase the amount of absorption or even produce gain, depending on the strength of the interaction.

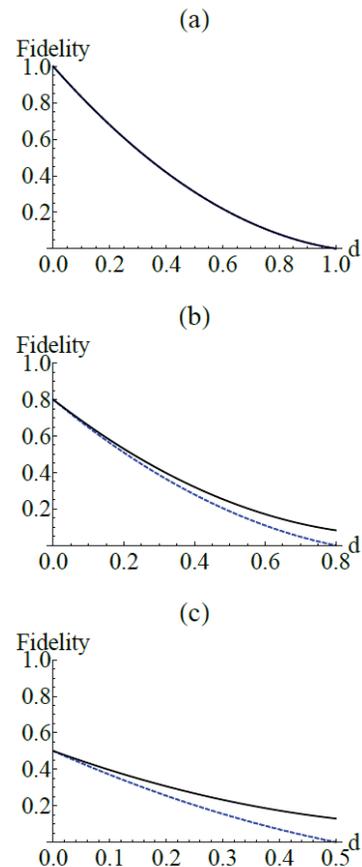

FIG. 7. Fidelity of the output stated plotted as a function of the dark count probability $d$ for several values of the loss probability $l$. The blue dashed curve shows the lower bound on the fidelity obtained by neglecting the contributions from the states $(0,2)$ and $(1,2)$ as described in the text. The black solid curve shows the actual fidelity without neglecting those terms. $\alpha = 2$ and $g = g_0(\alpha) = 1.154$ were chosen for the plots. (a) Loss probability $l = 0.0$. (b) $l = 0.2$ (c) $l = 0.5$.

The state produced by the post-selection process is orthogonal to a coherent state whose amplitude depends on the value of the gain. This can be understood as being due to cancellation between the two gain-dependent terms in Eq. (9), which produces an asymmetric dependence of the coefficient $c_n$ as a function of $n$ as illustrated in Fig. 2. A corresponding zero in the Q-function is apparent in Fig. 5.

This orthogonality may be a useful property when using these states as continuous variable qubits.

It can be seen from Eq. (10) that the probability of success becomes exponentially small for coherent state inputs with a large amplitude, since it becomes increasingly unlikely that one or more pairs of photons will not be emitted due to stimulated emission in the signal mode. This is one of the major open problems in quantum state engineering using conditional measurements. Nevertheless, this approach may have useful applications for moderate values of the gain and input coherent state amplitudes. The fidelity of the output state is primarily limited by photon loss or detector efficiency, but reasonably high values of the fidelity should be achievable.

Our analysis provides an interesting example of the variety of quantum states that can be obtained by varying the gain in a post-selected optical parametric amplifier. In addition, this approach may have practical applications for moderate values of $\alpha$, since the gain and the output state can be continuously varied by adjusting the intensity of the pump beam. Two amplifiers can also be used in a somewhat similar technique to create entangled macroscopic states, as will be discussed in a separate paper.

**Acknowledgements**

We would like to acknowledge valuable discussions with Richard Brewster, Ian Nodurft, Cory Nunn, and Todd Pittman. This work was supported in part by the National Science Foundation under grant number PHY-1802472.

**References**

[1]  G. S. Agarwal and K. Tara, Physical Review A (Atomic, Molecular, and Optical Physics) **43**, 492 (1991).
[2]  A. Zavatta, S. Viciani, and M. Bellini, Science **306**, 660 (2004).
[3]  V. Parigi, A. Zavatta, K. Myungshik, and M. Bellini, Science **317**, 1890 (2007).
[4]  M. Barbieri, N. Spagnolo, M. G. Genoni, F. Ferreyrol, R. Blandino, M. G. A. Paris, P. Grangier, and R. Tualle-Brouri, Physical Review A (Atomic, Molecular, and Optical Physics) **82**, 063833 (2010).
[5]  S. N. Filippov, V. I. Man'ko, A. S. Coelho, A. Zavatta, and M. Bellini, Physica Scripta T **2013**, 014025 (2013).
[6]  M. Boiteux and A. Levelut, Journal of Physics A (Mathematical and General) **6**, 589 (1973).
[7]  F. A. M. de Oliveira, M. S. Kim, P. L. Knight, and V. Buzek, Physical Review A (Atomic, Molecular, and Optical Physics) **41**, 2645 (1990).
[8]  A. I. Lvovsky and S. A. Babichev, Physical Review A (Atomic, Molecular, and Optical Physics) **66**, 011801/1 (2002).
[9]  S. L. Braunstein and P. van Loock, Reviews of Modern Physics **77**, 513 (2005).
[10]  M. Miranda and D. Mundarain, Quantum Information Processing **16**, 298 (2017).
[11]  W. Dong, L. Mo, Z. Feng, Y. Zhen-Qiang, C. Wei, H. Zheng-Fu, G. Guang-Can, and W. Qin, Physical Review A (Atomic, Molecular, and Optical Physics) **90**, 062315 (2014).
[12]  K. P. Seshadreesan, J. P. Olson, K. R. Motes, P. P. Rohde, and J. P. Dowling, Physical Review A (Atomic, Molecular, and Optical Physics) **91**, 022334 (2015).
[13]  C. H. Bennett, G. Brassard, C. Crepeau, R. Jozsa, A. Peres, and W. K. Wootters, Physical Review Letters **70**, 1895 (1993).
[14]  H. Jeong and M. S. Kim, Physical Review A (Atomic, Molecular, and Optical Physics) **65**, 042305/1 (2002).
[15]  S. A. Podoshvedov, Physical Review A (Atomic, Molecular, and Optical Physics) **79**, 012319 (2009).
[16]  J. Sperling, W. Vogel, and G. S. Agarwal, Physical Review A **89**, 043829 (2014).
[17]  M. Dakna, L. Knoll, and D. G. Welsch, Optics Communications **145**, 309 (1998).
[18]  E. Bimbard, N. Jain, A. MacRae, and A. I. Lvovsky, Nature Photonics **4**, 243 (2010).
[19]  D. T. Pegg, L. S. Phillips, and S. M. Barnett, Physical Review Letters **81**, 1604 (1998).
[20]  M. Ban, Optics Communications **143**, 225 (1997).
[21]  S. M. Barnett, D. T. Pegg, and J. Jeffers, Optics Communications **172**, 55 (1999).
[22]  R. A. Brewster, I. C. Nodurft, T. B. Pittman, and J. D. Franson, Physical Review A **96**, 042307 (2017).
[23]  S. Sivakumar, Physical Review A (Atomic, Molecular, and Optical Physics) **83**, 035802 (2011).
[24]  A. Luis and J. Peřina, Physical Review A **53**, 1886 (1996).
[25]  C. Brif and A. Mann, Quantum and Semiclassical Optics: Journal of the European Optical Society Part B **9**, 899 (1997).
[26]  B. L. Schumaker and C. M. Caves, Physical Review A (General Physics) **31**, 3093 (1985).
[27]  C. M. Caves, J. Combes, J. Zhang, and S. Pandey, Physical Review A (Atomic, Molecular, and Optical Physics) **86**, 063802 (2012).
[28]  K. Husimi, Proceedings of the Physico-Mathematical Society of Japan. 3rd Series **22**, 264 (1940).
[29]  Y. Kano, Journal of Mathematical Physics **6**, 1913 (1965).
[30]  M. D. Eisaman, J. Fan, A. Migdall, and S. V. Polyakov, Rev. Sci. Instrum. **82**, 071101 (2011).
[31]  I. C. Nodurft, R. A. Brewster, T. B. Pittman, and J. D. Franson, Physical Review A **100**, 013850 (2019).

## Appendix

The mean and variance of the photon number were briefly discussed in the main text. The purpose of this appendix is to discuss some of their properties in more detail.

For convenience, we rewrite the output state from Eq. (21) in the form

$$|\psi\rangle = C_0 \left|\frac{\alpha}{g}\right\rangle + C_1 \left|\frac{\alpha}{g},1\right\rangle, \tag{A1}$$

where

$$C_0 \equiv \frac{1}{\sqrt{N}}\left(\frac{1}{g^2} - \left|\frac{\alpha}{g}\right|^2 G^2\right) \tag{A2}$$

and

$$C_1 \equiv -\frac{1}{\sqrt{N}}\frac{\alpha}{g}G^2. \tag{A3}$$

Using these definitions, the average photon number can be shown to be

$$\langle \hat{n} \rangle = |C_0|^2 \left\langle \frac{\alpha}{g}\left|\hat{n}\right|\frac{\alpha}{g}\right\rangle + |C_1|^2 \left\langle \frac{\alpha}{g},1\left|\hat{n}\right|\frac{\alpha}{g},1\right\rangle$$
$$+ 2\operatorname{Re}\left(C_0^* C_1 \left\langle \frac{\alpha}{g}\left|\hat{n}\right|\frac{\alpha}{g},1\right\rangle\right). \tag{A4}$$

This can be simplified further by using $\hat{n} = \hat{a}^\dagger \hat{a}$ to give

$$\langle \hat{n} \rangle = |C_0|^2 \left|\frac{\alpha}{g}\right|^2 + |C_1|^2 \left(\left|\frac{\alpha}{g}\right|^2 + 1\right)$$
$$- \frac{2}{N}\left|\frac{\alpha}{g}\right|^2 \left(\frac{1}{g^2} - \left|\frac{\alpha}{g}\right|^2 G^2\right). \tag{A5}$$

Fig. 8 shows the behavior of the average photon number as a function of the amplifier gain. We observe that the average photon number initially decreases as we increase the gain before increasing to a maximum value at a gain of approximately $g_0$. This behavior can also be seen in Fig. 5 where the peak value shifts closer to the origin between Fig. 5 (a) and (b) whereas the peak shifts away from the origin between Fig. 5 (b) and (d).

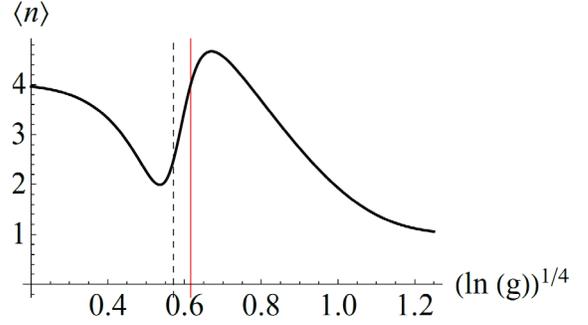

FIG. 8. Average photon number in the output signal mode as a function of the amplifier gain $g$, for an incident coherent state amplitude of $\alpha = 2$. The solid red vertical line corresponds to a gain of $g = g_0$ while the black dashed vertical line corresponds to a gain of $g = g_1$.

The second moment of photon number in the final state can also be calculated using a similar procedure to give

$$\langle \hat{n}^2 \rangle = |C_0|^2 \left(\left|\frac{\alpha}{g}\right|^2 + \left|\frac{\alpha}{g}\right|^4\right)$$
$$+ |C_1|^2 \left(3\left|\frac{\alpha}{g}\right|^2 + \left(\left|\frac{\alpha}{g}\right|^2 + 1\right)^2\right) \tag{A6}$$
$$- \frac{2}{N}\left|\frac{\alpha}{g}\right|^2 \left(\frac{1}{g^2} - \left|\frac{\alpha}{g}\right|^2 G^2\right)\left(2\left|\frac{\alpha}{g}\right|^2 + 1\right),$$

from which the variance in the photon number can be calculated using

$$\operatorname{Var}(n) = \langle \hat{n}^2 \rangle - \langle \hat{n} \rangle^2. \tag{A7}$$

A plot of the variance is shown in Fig. 9 for an input coherent state amplitude of $\alpha = 2$. It can be seen that the variance increases rapidly to a maximum value at a gain of approximately $g_0$. This feature could be useful in experiments for determining if the output state is close to the displaced single photon state. Another interesting feature that can be seen in these plots is that the average photon number approaches unity in the limit of high gain while the variance vanishes, suggesting that a single photon state is produced in that limit.

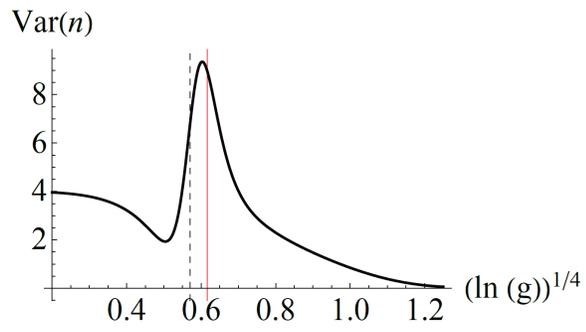

FIG. 9. Variance in the photon number of the output signal mode as a function of amplifier gain for the case in which the input coherent state has an amplitude of $\alpha = 2$. As in Fig. 8, the solid red vertical line corresponds to a gain of $g = g_0$ while the black dashed vertical line corresponds to a gain of $g = g_1$.